\documentclass[twocolumn,amsmath,amssymb,prd,superscriptaddress,preprintnumbers,letterpaper,showpacs,floatfix]{revtex4}

\usepackage{graphicx}
\usepackage{latexsym}
\usepackage{bm}
\usepackage{longtable}

\newcommand{\beq}[1]{\begin{equation}\label{#1}}
\newcommand{\eeq}{\end{equation}}
\newcommand{\bea}[1]{\begin{eqnarray} \label{#1}}
\newcommand{\eea}{\end{eqnarray}}
\newcommand{\ba}{\begin{array}}
\newcommand{\ea}{\end{array}}

\newcommand{\rf}[1]{(\ref{#1})}

\newcommand{\lsim}{\mathrel{\vcenter{\hbox{$<$}\nointerlineskip\hbox{$\sim$}}}}
\newcommand{\gsim}{\mathrel{\vcenter{\hbox{$>$}\nointerlineskip\hbox{$\sim$}}}}

\newcommand{\half}{\frac{1}{2}}
\newcommand{\third}{\frac{1}{3}}
\newcommand{\quarter}{\frac{1}{4}}

\newcommand{\rarr}{\rightarrow}
\newcommand{\lrarr}{\leftrightarrow}

\def\fr#1#2{{{#1} \over {#2}}}

\newcommand{\dmsq}{\delta m^2}

\def\U2{\underline{U\hspace{-.9mm}}\hspace{.9mm}}
\def\P{\underline{P\hspace{-.7mm}}\hspace{.7mm}}

\def\Ena{E_{N\!A}}
\def\bbar{\bar b}

\def\nue{{\nu_{e}}}
\def\numu{{\nu_{\mu}}}
\def\nutau{{\nu_{\tau}}}
\def\nuebar{{\bar \nu}_e}
\def\numubar{{\bar \nu}_\mu}
\def\nutaubar{{\bar \nu}_\tau}
\newcommand{\nualpha}{\nu_\alpha}

\newcommand{\ketbra}{\rangle\langle}
\def\A2N{adiabatic-to-nonadiabatic}
\newcommand{\trx}{\theta_{13}}
\newcommand{\tatmo}{\theta_{32}}
\newcommand{\tsun}{\theta_{12}}
\newcommand{\xinu}{\xi_\nu}
\newcommand{\we}{w_e}
\newcommand{\wmu}{w_\mu}
\newcommand{\wtau}{w_\tau}

\newcommand{\Ehalf}{E_{1/2}}
\newcommand{\tildeG}{{\tilde\Gamma}}
\newcommand{\sgn}{{\rm sign}}

\begin{document}

\title{Flavor sensitivity to $\bm{\theta_{13}}$ and sign($\bm{\delta m^2_{32}}$) 
for neutrinos from solar WIMP annihilation}

\author{Ralf Lehnert}
\email[Electronic mail: ]{ralf.lehnert@nucleares.unam.mx}
\affiliation{Instituto de Ciencias Nucleares,
Universidad Nacional Aut\'onoma de M\'exico,
A.~Postal 70-543, 04510 M\'exico D.F., Mexico}

\author{Thomas J.\ Weiler}
\email[Electronic mail: ]{tom.weiler@vanderbilt.edu}
\affiliation{Department of Physics and Astronomy,
Vanderbilt University, Nashville, TN 37235, USA}

\date{February 11, 2010}

\begin{abstract} 
The effect of the higher-energy 2nd resonance 
and the associated adiabatic-to-nonadiabatic 
transition on neutrino propagation in solar matter is presented. 
For WIMP-annihilation neutrinos 
injected with energies in the ``sweet region'' between $300\,$MeV and $10\,$GeV
at the Sun's center, 
a significant and revealing dependence
on the neutrino mass hierarchy and the mixing angle $\theta_{13}$ 
down to $0.5^\circ$ is found 
in the flavor ratios arriving at Earth.
In addition, 
the amplification of flavor ratios in the sweet region 
allows a better discrimination 
among possible annihilation modes of the solar dark matter.  
Under mild assumptions on WIMP properties, 
it is estimated that 200 neutrino events in the sweet region 
would be required for inferences of $\theta_{13}$, 
the mass hierarchy, 
and the dominant WIMP annihilation mode.
Future large-volume, low-energy neutrino detectors 
are likely needed 
if the measurement is to be made.
\end{abstract}


\maketitle

\section{Introduction}
\label{sec:intro} 
The annihilation of weakly interacting massive particles (WIMPs), 
trapped in the Sun's core, 
is expected to produce fluxes of neutrinos and antineutrinos of all active flavors 
at energies well above the MeV scale of solar-fusion $\nu_e$'s. 
These WIMP-annihilation neutrinos encounter
the 2nd matter resonance at higher energy 
$E^h_R  \simeq|\delta m^2_{32}|\cos 2\theta_{13}/[2\, V_e(0)]
\sim 0.2\,$GeV.
We find that a small leptonic mixing angle $\theta_{13}$ 
will inject a tell-tale signal in the flavor spectrum at Earth 
in the energy region $0.3$--$10\,$GeV,  
much as the 1st resonance at lower energy 
$E^l_R\simeq\delta m^2_{21}\cos2\theta_{12}/[2\, V_e(0) \cos^2\theta_{13}]
\sim 1.8\,{\rm MeV}$
is expected to do 
in future measurements of the solar-fusion $\nu_e$ spectrum. 
This signal implies an experimental sensitivity to $\theta_{13}$,
presently constrained by CHOOZ data to be below $12^\circ$,
down to about half a degree.
In the above expressions, 
$\delta m^2_{jk}\equiv m^2_j -m^2_k$,
and $V_e(0)\sim 7\times 10^{-12}\,$eV is the matter potential 
resulting from the electron density at the Sun's core; 
the superscripts $h$ and $l$ denote the higher-energy 2nd resonance 
and the lower-energy 1st resonance, 
respectively. 
The signal comes from the adiabatic-to-nonadiabatic transition 
at the 2nd resonance.
This effect is the higher-energy boundary 
of the ``bathtub'' spectral shape, 
well-known and well-described for the 1st solar resonance, 
and elucidated for the 2nd solar resonance 
in Figs.~3 and 4 of Ref.~\cite{lw07}
as well as herein.

The Borexino experiment~\cite{Borexino} in progress 
and the SNO+ experiment~\cite{SNO+}
in the construction stage 
are likely to measure the 
$^7$Be and $pep$ monochromatic solar neutrinos 
at energies of $0.861\,$MeV and $1.442\,$MeV, 
respectively.
These measurements should reveal 
the lower-energy boundary of the bathtub profile from the 1st resonance.
The adiabatic-to-nonadiabatic transition at the 2nd resonance 
offers a potentially much more striking signal 
than the lower-energy resonance,
as we will show in the present work.
The heuristic reason is that 
MSW resonant enhancement of the small $\trx$ 
to the MSW resonant value of $45^\circ$
is a much larger effect than 
the enhancement of $\tsun\sim 32^\circ$ 
to the resonant $45^\circ$ value.
In fact, 
as we will demonstrate, 
the adiabatic-to-nonadiabatic transition at the 2nd resonance 
is extremely sensitive to the value of small $\theta_{13}$:
the resonance is completely nonadiabatic and ignorable for $\theta_{13}=0$,
whereas the adiabatic component is significantly amplified for nonzero $\theta_{13}$,
even as small as $0.5^\circ$.
In addition, 
the 2nd resonance occurs in the neutrino sector 
if the neutrino mass hierarchy is normal
(i.e., $\delta m^2_{32}>0$),
but in the antineutrino sector 
if the mass hierarchy is inverted 
(i.e., $\delta m^2_{32}<0$).
WIMP annihilation in the Sun 
is expected to produce neutrinos and antineutrinos in equal numbers.
The different absorption and scattering cross-sections of neutrinos versus antineutrinos 
in terrestrial detectors 
then allows possible discrimination between the two.
It thus becomes apparent that 
the 2nd resonance provides the potential to reveal 
not only the value of $\theta_{13}$,
but also the neutrino mass hierarchy 
codified as $\sgn (\delta m^2_{32})$.

As in the case of solar-fusion neutrinos, 
the total neutrino spectrum remains unaffected 
by the resonance structure~\cite{comment}.
It is rather the distribution of the neutrinos over the three flavors 
that depends on the resonance physics. 
For this reason, 
we focus our attention 
on the individual flavor spectra 
(three flavors each for neutrinos and antineutrinos) 
that experiments at Earth can, 
in principle, 
detect.

While much theoretical effort has gone into 
studies of ``indirect detection'' of solar WIMPs 
by identification of their high-energy neutrino flux 
in large terrestrial detectors~\cite{solarWIMPS},
little has been done~\cite{lw07,LearnedKumar} 
to elucidate the possibilities at lower energies, $E \alt 10\,$GeV.
The reasons are clear:  
most of the neutrino flux from WIMP annihilation 
is expected to populate the higher-energy region,
the atmospheric background falls as $\sim E^{-3}$ above $10\,$GeV~\cite{Bartol},
and the detection cross section for a neutrino 
grows as $\sim E$ above GeV neutrino energies.
However, 
the substantial discovery potential that 
hides in the solar WIMP-annihilation neutrino data below $10\,$GeV 
compels us to promote the associated lower-energy physics.  
In the next section, 
we present an overview of this rich neutrino physics 
at lower energies.
Details, 
drawn mainly from Ref.~\cite{lw07}, 
are presented in subsequent sections.

%%%%%%%%%%%%%%%%%%%%%%%%%%%%%%%%%%%%%%%%%%%%%%%%%%%%%%%%%%%%%%%%%%%%%%%%%%%%%%%%%%%%

\section{Preliminaries and overview}
\label{sec:prelim}
A statistical average over the oscillation phase $\phi=\delta m^2_{jk}\,L/2E$ 
effectively results from 
the uncertainties in the baseline $L$ and energy $E$ of solar neutrinos. 
The production flavor ratios $w_{\alpha}(E)$ in the Sun's core 
and the terrestrial flavor ratios $W_{\alpha}(E)$ for neutrinos 
are then related by~\cite{lw07}:
\beq{terrestrial_flavor_ratios} 
\left(
\ba{l}
W_e\\
W_\mu\\
W_\tau
\ea
\right)
= \U2\,\P\,\U2_{\hspace{-.3mm}m}^T (r=0)
\left(
\ba{l}
\we\\
\wmu\\
\wtau
\ea
\right)\,.
\eeq 
Here, 
$(\U2_{\hspace{-.3mm}m})_{\alpha j}\equiv |(U_m)_{\alpha j}|^2$ 
is a classical probability matrix, 
constructed from the mixing matrix in matter $U_m$. 
The transposed matrix $\U2_{\hspace{-.3mm}m}^T (r)$ 
transforms the production flavor fluxes  
to the propagating mass-state fluxes. 
The argument $r=0$ in Eq.~\rf{terrestrial_flavor_ratios} reminds us that 
WIMP annihilation occurs in the solar core, 
and so it is the matter density at the Sun's center that 
determines $\U2_{\hspace{-.3mm}m}^T (0)$.

Possible nonadiabatic transitions 
between the effective-mass states 
at the 1st and 2nd resonances 
and other solar-matter effects
are described by 
the level-crossing probability matrix $\P$~\cite{Parke,KuoPanta,alt_matter}.
Finally, the matrix $\U2_{\alpha j}\equiv |U_{\alpha j}|^2$, 
determined from the usual vacuum mixing matrix $U$,
transforms the mass states back to detectable flavor fluxes at Earth.
Explicit expressions for $U_m$ and $\P$ 
in terms of neutrino energy $E$, 
solar-model coefficients, 
and neutrino parameters, 
as well as the analogous result for antineutrinos 
can be found in Ref.~\cite{lw07}, 
along with many other relevant details.
Some related references that 
study flavor issues for solar neutrinos arising from WIMP annihilation 
are listed in~Ref.~\cite{related}.

Inspection of Eq.~\rf{terrestrial_flavor_ratios} reveals that 
structures in the flavor spectra can potentially arise from 
three energy-dependent sources: 
the initial flavor ratios $w_\alpha (E)$, 
the mixing matrix in matter $\U2_{\hspace{-.3mm}m}(r=0,E)$, 
and the jump-probability matrix $\P(E)$. 
The vacuum mixing matrix $U$,
and thus $\U2$, 
are each independent of the neutrino energy, 
and so $\U2$ does not contribute to energy-dependent features 
in the terrestrial flavor spectra.

\begin{figure}
\begin{center}
\includegraphics[width=0.95\hsize]{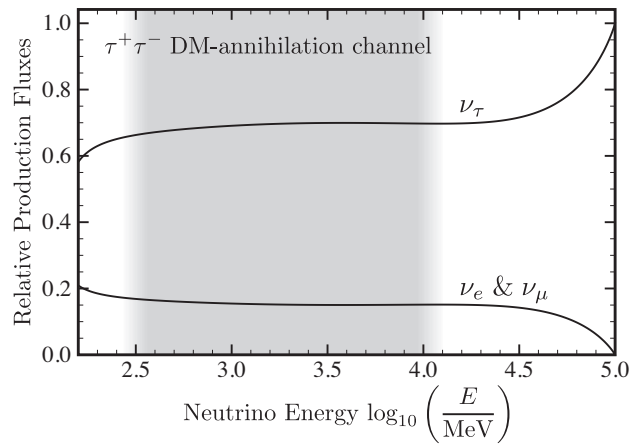}
\end{center}
\vskip-15pt
\caption{
Shown are the relative neutrino flavor spectra 
resulting from annihilation of WIMPs with $100\,$GeV mass 
into predominantly $\tau^+\tau^-$. 
This plot is based on the numerical results 
obtained by Cirelli {\em et al.}~\cite{Cirelli05}.
The shaded area represents the ``sweet region'' defined in the text.
Note the nearly constant values of the relative fluxes
inside this region.
}
\label{fig0} 
\end{figure}

The first of the above sources for features in the terrestrial flavor spectra, 
namely $w_\alpha$, 
is WIMP-model dependent.
A description of this energy dependence 
requires knowledge of the WIMP$\to\nu_{\alpha}$ annihilation chains.
Three possible decay chains are commonly invoked.
They are neutrino production 
via decay of the intermediate states $W^+ W^-$, $b\,\bbar$, and $\tau^+\tau^-$.
The latter two chains lead to softer neutrino spectra 
than does the former.
Calculations~\cite{Cirelli05} reveal that 
for the $b\,\bbar$ and $\tau^+\tau^-$ annihilation chains, 
the neutrino flavor ratios at production 
are slowly varying in the low-energy region~\cite{fn}.
For the $W^+W^-$ chain, 
these flavor ratios are more energy-dependent~\cite{WWnot}.
We show the nearly energy-independent neutrino flavor spectra 
for $M_{\rm WIMP}=100\,$GeV annihilation
via the $\tau^+\tau^-$ mode in Fig.~\ref{fig0}.
For a $100\,$GeV WIMP mass, 
the region of nearly constant flavor ratios 
lies below $20$--$30\,$GeV~\cite{absorpnote}.
For specificity in the rest of this paper, 
we will continue to focus on the $b\,\bbar$ and (especially) $\tau^+\tau^-$ modes,
as well as on a WIMP mass of order $100\,$GeV.
If the dominant annihilation mode of the WIMP is $W^+ W^-$ 
rather than $\tau^+\tau^-$ and/or $b\,\bbar$,
then we expect that 
the resonant flavor change 
that we describe below 
can still be extracted from data,
but further efforts would be required to isolate the resonant features 
from the non-resonant energy-varying flavor ratios.

%%%%%%%%%%%%%%%%%%%%%%%%%%%%%%%%%%%%%%%%%%%%%%%%%%%%%%%%%%%%%%%%%%%%
%%%%%%%%%%%%%%%%%%%%%%%%%%%%%%%%%%%%%%%%%%%%%%%%%%%%%%%%%%%%%%%%%%%%

The second potential source of energy dependence, 
the matrix $\U2_{\hspace{-.3mm}m}$, 
can also be tamed: 
for neutrino energies sufficiently above the 2nd resonance, 
$\U2_{\hspace{-.3mm}m}$ approaches an $E$-independent constant matrix 
(see e.g., Ref.~\cite{lw07}). 
For this reason, 
we will consider only energies well above $E_R^h \sim 0.3\,$GeV, 
so that $\U2_{\hspace{-.3mm}m}$ is effectively constant. 

Some energy dependence will also be introduced in the flavor ratios  
by the adiabatic-to-nonadiabatic transition of the 1st resonance.
This occurs  at energies at and above $\Ena^l\sim 10\,$GeV~\cite{lw07}.
As can be seen in Figs.~3 and~4 of Ref.~\cite{lw07}, 
the effects of the 1st resonance are mild;
to ignore these effects would introduce little uncertainty into our calculation,
and the effects could be included in a more complete analysis.
However, 
we will simply restrict our study to energies below $\Ena^l$
to shield our analysis from ``contamination'' due to the 1st resonance. 
Thus, 
we arrive at, and define, a ``sweet region'' in energy for our analysis:
\beq{sweet_region}
0.3\,{\rm GeV} \le E \le 10\,{\rm GeV}\quad \textrm{``SWEET REGION.''}
\eeq
The results presented below 
will hold within this special region of energy.

The remaining and most important source of structure 
in the flavor spectra 
arises from the neutrino level-crossing probabilities at the resonances, 
which are described by the jump matrix $\P$.
The transition from the adiabatic (i.e., no level-crossing) 
to the nonadiabatic (i.e., complete level-crossing) regime at high energies 
can leave a dramatic imprint in the terrestrial flavor-flux ratios. 

In Ref.~\cite{lw07}, 
the energy $\Ena ^h$ characterizing the onset 
of this adiabatic-to-nonadiabatic transition  
was defined implicitly by setting $P_c^h(\Ena ^h)$
in the crossing-probability matrix 
to $e^{-3}\simeq5\%$.
Numerically, 
this onset energy is 
$\Ena ^h\simeq 750\,\sin^2\theta_{13}\,$GeV.
(An expression for the crossing probability is given below in Eq.~\rf{crossprob}.)
The 2nd resonance occurs in the neutrino sector for the normal hierarchy 
with $\delta m^2_{32} >0$,
and in the antineutrino sector for the inverted hierarchy with $\delta m^2_{32} <0$.

The above discussion shows that 
we may identify any observable energy dependence of the flavor spectra
in the sweet region 
as due to the adiabatic-to-nonadiabatic transition at $\Ena^h$. 
In turn, 
the energy dependence in this sweet region 
will implicate the value of a small but nonzero $\theta_{13}$ 
as well as the neutrino mass hierarchy.
In particular, at small $\theta_{13}$ 
the nonadiabatic jump probability is $\exp(-\Gamma\,\theta_{13}^2)$.
The ``adiabaticity parameter'' $\Gamma$ 
can be written in terms of the length scale  $\lambda_\odot = |\frac{d}{dr} \ln N_e|^{-1}$ 
for the change in the electron density $N_e$
at the resonance region 
(corresponding to one $e$-folding in density for an exponential profile like that of the Sun),
and the neutrino oscillation length in vacuum $\lambda_v=4\pi\,E/\delta m^2$  
as $\Gamma=4\pi^2\,\lambda_\odot/\lambda_v$.
The important feature for our purposes is that 
$\Gamma$ scales as $1/E$ 
and is therefore large, 
roughly $300$--$10\,000$, 
across the low energies of our sweet region.
It is this large size of $\Gamma$ that 
produces an observable signal 
even for extremely small $\theta_{13}$.
Qualitatively, 
we expect sensitivity to $\trx$ in the sweet region 
for $\trx$ as small as $1/\sqrt{\Gamma}\agt 10^{-2}\sim 0.5^\circ$.
In Sec.~\ref{sec:theory}, 
we show quantitatively that this is indeed the case.

There is also a terrestrial source of energy dependence in the flavor ratios.
This arises from matter effects for neutrinos and antineutrinos traversing the Earth.
The effect has been well worked out in publications~\cite{EarthMatter};
it is somewhat complicated 
(best addressed with a numerical code), 
and we will not include it in this paper.
We do give here some general remarks~\cite{lw07} 
about the Earth matter effect 
that are relevant for our present purposes.
The resonant energies in Earth 
related to the solar scale $\dmsq_{21}$ 
are below the sweet region
(at $\sim 25\,$MeV and $\sim 100\,$MeV 
for the Earth's core and mantle, 
respectively),
while the  resonant energies in Earth 
related to the atmospheric-scale $\dmsq_{32}$ 
fall right in the sweet region,
at $\sim 2\,$GeV and $\sim10\,$GeV for the Earth's core and mantle, 
respectively.
As with the solar resonance, 
for a normal mass hierarchy 
the relevant resonance in Earth 
occurs in the neutrino but not the antineutrino sector.
And again as with the solar resonance, 
for an inverted mass hierarchy the relevant resonance in Earth 
occurs in the antineutrino but not the neutrino sector.
The resonant amplification can be quite large, 
but the effect can be mitigated by 
a resonant oscillation wavelength scaling like $1/\theta_{13}$. 
The requirement that at least a quarter of the oscillation length 
must lie within the Earth to ``feel'' the matter effects
leads to a nadir-dependent condition on $\theta_{13}$.
For a neutrino with zero nadir angle, 
$\theta_{13}\agt 0.5^\circ (E/{\rm GeV})$ is required to feel the matter;
for other nadir angles, 
larger values of $\theta_{13}$ are required to feel the matter.
In a more complete analysis involving real data,
terrestrial-matter effects should then no longer be ignored.

It is worth remarking here 
what would change 
if the WIMP mass were, 
say, 
a TeV rather than 
the $100\,$GeV we have assumed~\cite{fn}.  
Surely, 
the available phase space for the produced neutrinos is increased.
Moreover, 
in Ref.~\cite{Cirelli05} it is shown that 
the upper end of the neutrino energy spectrum scales linearly with the WIMP mass,
tending to enlarge the energy region 
in which the flavor ratios are constant.
However, 
the sweet region remains between $0.3$ and $10\,$GeV 
because it is determined solely by 
the effects of solar matter on neutrino propagation.
It follows that 
the phase space of the sweet region 
relative to the enlarged total phase space
now represents a smaller fraction, 
so that less sweet-region neutrinos are available.
More importantly, 
according to Eq.~\rf{capturerate}
the expected neutrino flux from WIMP annihilation in the Sun 
scales as $M_{\rm WIMP}^{-2}$, 
and so would be down by an additional factor of 100 for annihilation of TeV WIMPs 
compared to $100\,$GeV WIMPs.

One must inevitably ask 
what event rate at Earth 
might be expected from solar WIMP annihilation to (anti)neutrinos 
in the sweet region,
and what rate is needed to observe structure in the terrestrial flavor spectra.
We attend to first of these questions in the next section, 
and discuss the requirement for statistical significance 
in our final Sec.~\ref{sec:discon}.

%%%%%%%%%%%%%%%%%%%%%%%%%%%%%%%%%%%%%%%%%%%%%%%%%%%%%%%%%%%%%%%%%%%%%%%%%%%%%%%%%%%%

\section{(Anti)Neutrino flux from solar WIMP annihilation }
\label{sec:flux}
Theory suggests that 
the age of the Sun exceeds the equilibration time 
between solar capture of WIMPs and their subsequent annihilation in the Sun~\cite{gould}.
Consequently, 
the WIMP annihilation rate is given by 
half of the WIMP capture rate,
where the ``half'' just reflects the fact that 
it takes two captures to enable one two-body annihilation.
The WIMP capture rate by the Sun is given in Ref.~\cite{gould}:
\beq{capturerate}
C_\odot \simeq 1.0\times 10^{25}\,{\cal F}\;{\rm s}^{-1}\,,
\eeq
where
\beq{test1}
{}\hspace{-1mm}{\cal F}=
\left(\!\frac{\rho_{\rm WIMP}}{0.3\,{\rm \frac{GeV}{cm^3}}}\!\right)\!\!
\left(\!\frac{270\,{\rm \frac{km}{s}}}{v_{\rm WIMP}}\!\right)\!\!
\left(\frac{\sigma_{\rm WIMP}}{3\!\times\! 10^{-38}{\rm cm}^2}\right)\!\!
\left(\frac{100\,{\rm GeV}}{M_{\rm WIMP}}\right)^{\!2}
\eeq
is a fiducial factor. 
Here, 
$\sigma_{\rm WIMP}$ and $M_{\rm WIMP}$ 
are the unknown WIMP--nucleon scattering cross section and mass,
while $\rho_{\rm WIMP}$ and $v_{\rm WIMP}$ 
denote the density and rms velocity of the local WIMP population.
Each parenthetical fraction displays typical values, 
except that the fiducial value shown for $\sigma_{\rm WIMP}$
is the present upper limit for the spin-dependent cross section~\cite{SK04,IC22-09};
it must therefore be viewed as optimistic.
The inverse quadratic dependence of the rate on the WIMP mass 
is easily understood:
one factor arise from the conversion of WIMP mass density $\rho_{\rm WIMP}$ 
to number density,
and the other factor is kinematic, 
reducing the capture efficiency when 
the beam and target masses are mismatched.
The fiducial values for today's local WIMP density and rms velocity 
serve as numerical guidelines,
but in fact the integrated history of WIMP capture by the Sun 
over large look-back times 
may yield values that 
do not adhere to these guidelines.

The WIMP--nucleon cross section in Eq.~\rf{test1} above  
is the sum of a spin-independent cross section and a spin-dependent cross section,
each averaged over the target matter in the Sun's core.
Experiments on Earth that search directly for WIMPs typically 
use target materials with large nucleon number,
and so are better suited 
to limit (or detect) the spin-independent cross section~\cite{spin-indep}.
Direct-search limits on the spin-dependent cross section 
are five orders of magnitude weaker than 
the limits on the spin-independent cross section.
Since the solar material is mainly hydrogen, 
solar capture of WIMPs is very sensitive to the 
spin-dependent cross section. 

An experimental measurement of the solar neutrino flux from WIMP annihilation 
would bypass the theoretical uncertainties just listed.  
So far, 
experiments have yielded only upper limits on this flux 
inferred from final-state muons with energies above a GeV~\cite{SK04,IC22-09}.
It is interesting to note that 
the bounds on the spin-dependent WIMP cross section 
inferred from these experimental neutrino-flux constraints 
are stronger than the bounds 
coming from direct searches for WIMPs~\cite{directsearch}.

We continue by considering the mean multiplicity of neutrinos, 
$\xi_\nu$,
produced per annihilating WIMP (not WIMP pair), 
a quantity we anticipate to be of order unity.
We can then estimate the total neutrino flux at Earth~\cite{pi_note}:
\beq{Earthflux}
\int dE \,\frac{dN_\nu}{dE} \simeq \frac{C_\odot\,\xi_\nu}{4\pi\,({\rm AU})^2}
\sim \frac{ 1.1\times 10^5\,\xi_\nu\,{\cal F}}{\rm yr\cdot cm^2}\,.
\eeq
The background from ``atmospheric neutrinos,'' 
i.e., 
the neutrinos resulting from the decay of charged pions
produced by cosmic-ray interactions in our atmosphere, 
has been calculated by many groups with convergent results.
A typical energy spectrum in the GeV region can be found in Ref.~\cite{Bartol}.
It approximately obeys 
$1.2\times10^6\,\left(\frac{E}{\rm GeV}\right)^{-2.75}({\rm GeV\cdot yr\cdot cm^2})^{-1}$
for $(\numu+\numubar)$; 
the all-flavor flux would be about three times larger.
The integrated all-flavor flux above a GeV 
is then about $2\times 10^6/{\rm yr\ cm^2}$,
with the spectral-index factor nearly compensating the flavor factor.
This background flux is roughly twenty times 
the maximally allowed solar WIMP neutrino flux.
Moreover, 
the background flux has a flavor content that 
varies with direction on the sky.
The down-coming atmospheric neutrinos 
will show the $1:2$ $\nue$ to $\numu$ flavor ratio 
characteristic of the complete pion decay chain, 
as the muon decay length at lower energies is shorter than 
its atmospheric height,
and the neutrino pathlength is shorter than 
its vacuum oscillation length 
$L_{\rm osc}=4\pi\,E/\delta m^2\sim 1000\,(E/{\rm GeV})\,$km.
In contrast, 
the upcoming neutrinos will show a 1:1 flavor ratio, 
since half of the $\numu$'s will have oscillated into $\nutau$'s.
On the bright side, 
the solar fraction of solid angle on the sky is quite small, 
approximately $[\pi\,(R_\odot/{\rm AU})^2]/4\pi\sim5.4\times 10^{-6}$.
So one may hope that 
a cut favoring the direction to the Sun 
would greatly increase the signal-to-background ratio. 
However, 
the mean scattering angle for neutrinos below $10\,$GeV is large,
$\theta_{\rm scatt}\sim 20^\circ \sqrt{10\,{\rm GeV}/E}$.
The 24-hour angular modulation
expected in the solar signal will help reduce 
the unmodulated atmospheric background.
In addition, 
there has been some recent discussion of possible very large detectors~\cite{verylarge},
and some development 
in directional reconstruction of neutrinos 
at lower energies~\cite{Learned,Bernstein:2009ab}.
There have also been recent studies of using event topologies 
to achieve partial ``statistical'' separation of   
neutrino and antineutrino data samples~\cite{Schwetz}.

Multiplying the solar WIMP neutrino flux by 
(i) the neutrino--nucleon cross section, 
by (ii) the target number of nucleons given by
\beq{targetnumber}
N_N=6.0\times 10^{35}\left(\frac{M_{\rm target}}{{\rm megaton}}\right)\,,
\eeq
and by (iii) the fraction of incident neutrino flux in the sweet region 
$\Delta E(\textrm{SR})$ from 0.3 to $10\,$GeV
\beq{DeltaNnu}
f_\nu \equiv
\frac{\int_{\Delta E(SR)} d\ln E\,\frac{dN_\nu}{d\ln E}}
{\int d\ln E\,\frac{dN_\nu}{d\ln E}}\,,
\eeq
one obtains the event rate 
within the sweet energy region.
For a $100\,$GeV WIMP mass, 
inspection of the theoretical neutrino spectra in Ref.~\cite{Cirelli05}
suggests a value $\sim 20\%$ for $f_\nu$,
the fraction of neutrino flux in the sweet region between 0.3 and $10\,$GeV. 
As a fiducial event rate, 
we therefore take 
\beq{fidrate}
{}\hspace{-1.5mm}R=
140
\left(\frac{\sigma_{\nu N}(E)}{10^{-38}{\rm cm}^2}\right)\!
\left(\frac{M_{\rm target}}{{\rm megaton}}\right)\!
\left(\frac{f_\nu}{20\%}\right)\frac{\xinu\,{\cal F}}
{\rm yr}\,.
\eeq
For some perspective on the 140~events per year,
we may ask how many events per year 
are to be expected in a proton-decay experiment 
at a megaton detector.
The present limits on the lifetime of protons 
to decay to various modes are typically $10^{35}\,$yr. 
The nucleon number in a megaton is 
$10^{12}\,N_A = 6\times 10^{35}$. 
So the expected event rate is 
\beq{pdecay}
R_{p^+{\rm decay}} = 6\,\left( \frac{10^{35}\,{\rm yr}}{\tau_p} \right) 
\left( \frac{M_{\rm detector}}{\rm megaton} \right)\, {\rm yr}^{-1}\,.
\eeq
We see that 
the rates for detection of solar neutrinos in the sweet region from WIMP annihilation
and for detection for proton decay 
are comparable
(although the backgrounds are different).

%%%%%%%%%%%%%%%%%%%%%%%%%%%%%%%%%%%%%%%%%%%%%%%%%%%%%%%%%%%%%%%%%%%%%%%%%%

\section{Theoretical framework}
\label{sec:theory}
To proceed, 
we need explicit expressions 
for the neutrino flavor spectrum 
produced by WIMP annihilation in the Sun's core
in the range $0.3$--$10\,$GeV. 
We begin by parameterizing the relative production fluxes $w_{\alpha}$. 
In many models, 
the dominant WIMP-annihilation channels are 
$b\,\overline{b}$, $\tau^+\tau^-$, and $W^+W^-$. 
The neutrino spectrum from each channel 
has been calculated (see, e.g., Ref.~\cite{Cirelli05}), 
but the branching ratios to these channels 
depend on the specific model.  
Fortunately, 
WIMP decay obeys $w_e(E)=w_\mu(E)$ 
to a very good approximation~\cite{Cirelli05}.
Since the normalization $w_e+w_\mu+w_\tau=1$ holds
at any energy, 
the parametrization of the relative flavor spectra at production  
requires just one WIMP-model-dependent function $f_F(E)$.

We define $f_F$ operationally via 
$w_e=w_\mu=\frac{1}{3}-f_F$ and $w_\tau=\frac{1}{3}+2\,f_F$. 
Positivity of the $w_\alpha$ then implies 
the bound $-\frac{1}{6}\leq f_F\leq +\frac{1}{3}$. 
As defined, 
$f_F$ is the deficit of $w_e$ or $w_\mu$ from $\third$ at injection in the solar core.
Since maximal mixing effects symmetrization of $\numu$ and $\nutau$, 
it is also useful to view $f_F$ as the excess of $w_{\not \hspace{.2mm}e}\equiv w_\mu+w_\tau$ 
from $\frac{2}{3}$.
To order $f_F^2$, 
the ratios ($w_e/w_\tau$) and ($w_\mu/w_\tau$) at injection 
are $1-9\,f_F +36\,f_F^2$.

For comparative purposes later, 
it is also useful interpret $f_F$ in terms of the evolved flavor probabilities 
$W_\alpha$ in the {\em absence} of matter effects.
The flavor density matrix at injection, in the flavor basis, is given by
\beq{density-matrix1}
\rho_F=\third\,\openone+f_F\,\left[\,2\,|\nutau\ketbra\nutau|-|\numu\ketbra\numu|-|\nue\ketbra\nue|\,\right]\,.
\eeq
Using the tribimaximal mixing values to write this in the mass basis,
and invoking phase averaging to remove the off-diagonal elements,
one is left with just
\beq{density-matrix2}
\rho_F=\third\,\openone+\half\,f_F\,\left[\,|\nu_3\ketbra\nu_3|-|\nu_1\ketbra\nu_1|\,\right]\,.
\eeq
Then, 
the relative flavor probabilities 
$W_\alpha$ are just $\langle\,\nualpha\,|\,\rho_F\,|\,\nualpha\rangle$.
The results for vacuum transitions are 
$W_e={\overline{W}}_e=\third-\third\,f_F$, 
and $W_\mu=W_\tau={\overline{W}}_\mu={\overline{W}}_\tau= \third+\frac{1}{6}\,f_F$.
The vacuum value for the ratios $W_\mu/W_e$ and ${\overline{W}}_\mu/{\overline{W}}_e$ 
to order $f_F^2$ are $1+\frac{3}{2}\,f_F+f_F^2$.

Although actual WIMP properties are unknown, 
we may consider two of the aforementioned annihilation channels, 
$b\,\overline{b}$ and $\tau^+\tau^-$, 
as generic examples.  
We have argued that in our sweet region, 
the function $f_F(E)$ is reduced to an energy-independent number $f_F$ 
for the $\tau^+\tau^-$ and $b\,\bbar$ annihilation modes of the solar WIMPs.
Our fitted values for $f_F$ in these channels 
are $-0.09 \pm0.01$ and $+0.182\pm0.001$, respectively.

It is worth mentioning that 
in some WIMP models the annihilation to fermions 
proceeds through Higgs-like couplings.
We then expect the $b\,\bbar$ mode to dominate, 
but also a branching fraction to 
$\tau^+\tau^-$ given by $(m_\tau/m_b)^2/3\sim 5\%$, 
where the $\third$ reflects the $b\,\bbar$ mode's color factor.
In the sweet region, 
the $f_F$'s for the $b\,\bbar$ and $\tau^+\tau^-$ modes have opposite signs,
so the {\em average} $f_F$ for the fermionic channel 
is smaller in magnitude than $f_F(b\,\bbar)$ and $f_F(\tau^+\tau^-)$. 
However, 
the $b\,\bbar$ mode, 
and thus its concomitant $f_F$~value, 
dominate: 
an estimate employing the injection spectra 
given in Ref.~\cite{Cirelli05} 
indeed yields $f_F(\rm Higgs)\simeq-0.07\pm0.01$ 
in proximity to the $-0.09$ value for $b\,\bbar$ mode.
We remark that the neutrino spectrum from the $\tau^+\tau^-$ 
mode is harder than that 
from the  $b\,\bbar$ mode, and at energies above the sweet region, 
of no relevance to the present paper,
the $\tau^+\tau^-$ mode becomes increasingly important and eventually comes to dominate.
For the $W^+W^-$ mode, 
$f_F$ behaves quite differently:  
it rises nearly linearly with energy in the sweet region: 
$f_F(W^+W^-)=-0.029 + 0.014\,(E/\textrm{GeV})$. 
The values for the various $f_F$'s presented in this paragraph 
have been estimated 
using the results in Ref.~\cite{Cirelli05};  
they are also collected in Table~\ref{fF_table}.

\begin{table}
\caption{\label{fF_table} 
Values of $f_F=\third-w_e=\third-w_\mu=\half(w_\tau-\third)$ 
in the sweet region for popular WIMP annihilation modes.}
\renewcommand{\arraystretch}{1}
\vskip5pt
\begin{ruledtabular}
\begin{tabular}{cc}
WIMP annihilation mode & $f_F\ \ ({\rm sweet\ region})\rule[-3mm]{0mm}{1mm}$ \\ 
\hline
$\tau^+\tau^-\rule[+4mm]{0mm}{1mm}$ & \ \ $+0.182\pm0.001$ \\ 
$b\,\bbar\rule[+4mm]{0mm}{1mm}$\ \ \ & $-0.09 \pm0.01$ \\ 
Higgs-like & \ $-0.07\pm0.01\rule[+4mm]{0mm}{1mm}$ \\ 
$W^+W^-\rule[+4mm]{0mm}{1mm}$ & $-0.029 + 0.014\,(E/\textrm{GeV})$ \\ 
\end{tabular}
\end{ruledtabular}
\vskip-5pt
\end{table}

We note that 
for equal initial flavors ($f_F$=0), 
the resulting terrestrial flavor ratios are also democratic~\cite{lw07}, 
i.e., 
$W_e=W_\mu=W_\tau=1/3$. 
This is most easily seen by noting that 
the commutator in the density-matrix evolution equation
$i\,\frac{d\rho}{dt}=[ H,\,\rho ]$ vanishes for $\rho$ proportional to the identity matrix 
regardless of whether $H$ is the vacuum or matter Hamiltonian.
Thus, 
it is only the order 7--18\% $f_F$-value differences in the initial flavor spectrum 
that evolve nontrivially.
If flavor differences were not amplified by intervening matter effects, 
an event sample of about $(n/15\%)^2$ 
would be necessary to yield an $n$-sigma statistical inference of this difference.  
However, 
we will see shortly that 
matter effects, quadratically sensitive to $\theta_{13}$ 
and to the neutrino mass hierarchy, 
may amplify the flavor difference considerably.

\begin{table*}[t]
\caption{\label{table} Leading-order contributions 
to the components of the $A$ and $B$ flavor vectors in Eq.~\rf{general}.
}
\renewcommand{\arraystretch}{1}
\vskip5pt
\begin{ruledtabular}
\begin{tabular}{lccc}
\multicolumn{1}{c}{type\rule[-3mm]{0mm}{1mm}}
& $e$ flavor & $\mu$ flavor & $\tau$ flavor \\ 
\hline
%%%%%%%%%%%%%%%%%%%%%%%%%%%%%%%%%%%%%%%%%%%%%%%%%%%%%%%%%%%%%%%%%%%%%%%%%%%%%%%%%%%%%%
$A^{\rm NH}_\alpha\rule[+4mm]{0mm}{1mm}$ & $2+4\,\delta\theta_{23}$ & $-1-8\,\delta\theta_{23}$ & $-1+4\,\delta\theta_{23}$ \\
%%%
$B^{\rm NH}_\alpha$ & $-2-4\sqrt{2}\,\delta\theta_{12}+4\,\delta\theta_{23}$ & 
$1+\sqrt{8}\,\delta\theta_{12}+8\,\delta\theta_{23}+\sqrt{8}\,\theta_{13}\cos\delta$ & 
$1+\sqrt{8}\,\delta\theta_{12}-12\,\delta\theta_{23}-\sqrt{8}\,\theta_{13}\cos\delta$ \\
%%%%%%%%%%%%%%%%%%%%%%%%%%%%%%%%%%%%%%%%%%%%%%%%%%%%%%%%%%%%%%%%%%%%%%%%%%%%%%%%%%%%%%%
$\overline{A}{}^{\rm NH}_\alpha$ & $-2 + 4\sqrt{2}\,\delta\theta_{12} + 4\,\delta\theta_{23}$ & 
$1-\sqrt{8}\,\delta\theta_{12}-\sqrt{8}\,\theta_{13}\cos\delta$ & 
$1-\sqrt{8}\,\delta\theta_{12}+\sqrt{8}\,\theta_{13}\cos\delta-4\,\delta\theta_{23}$ \\
%%%
$\overline{B}{}^{\rm NH}_\alpha$ & $0$ & $0$ & $0$ \\
%%%%%%%%%%%%%%%%%%%%%%%%%%%%%%%%%%%%%%%%%%%%%%%%%%%%%%%%%%%%%%%%%%%%%%%%%%%%%%%%%%%%%%%
$A^{\rm IH}_\alpha$ & $-4\sqrt{2}\,\delta\theta_{12} + 8\,\delta\theta_{23}$ & 
$\sqrt{8}\,\delta\theta_{12}+\sqrt{8}\,\theta_{13}\cos\delta$ & 
$\sqrt{8}\,\delta\theta_{12}-\sqrt{8}\,\theta_{13}\cos\delta-8\,\delta\theta_{23}$ \\
%%%
$B^{\rm IH}_\alpha$ & $0$ & $0$ & $0$ \\
%%%%%%%%%%%%%%%%%%%%%%%%%%%%%%%%%%%%%%%%%%%%%%%%%%%%%%%%%%%%%%%%%%%%%%%%%%%%%%%%%%%%%%%
$\overline{A}{}^{\rm IH}_\alpha$ & $2-4\,\delta\theta_{23}$ & 
$-1-4\,\delta\theta_{23}$ & 
$-1+8\,\delta\theta_{23}$ \\
%%%
$\overline{B}{}^{\rm IH}_\alpha$\rule[-3mm]{0mm}{1mm} & 
$-4+4\sqrt{2}\,\delta\theta_{12}+8\,\delta\theta_{23}$ & 
$2-\sqrt{8}\,\delta\theta_{12}+4\,\delta\theta_{23}-\sqrt{8}\,\theta_{13}\cos\delta $ & 
$2-\sqrt{8}\,\delta\theta_{12}-12\,\delta\theta_{23}+\sqrt{8}\,\theta_{13}\cos\delta$ \\
\end{tabular}
\end{ruledtabular}
\vskip-5pt
\end{table*}

For further progress, 
we also need explicit expressions for $\U2$, $\P$, and $\U2_{\hspace{-.3mm}m}$
appearing in Eq.~\rf{terrestrial_flavor_ratios}. 
In the sweet region, 
the level-crossing probability at the lower resonance is zero, 
as explained above. 
The fact that Eq.~\rf{terrestrial_flavor_ratios} is linear in the level-crossing probability matrix 
$\P$ 
allows one to write a simple equation for the flavor evolution,
\beq{general}
W_\alpha=\fr{1}{3}+\fr{1}{4}\left[A_\alpha+B_\alpha\, P_c^h(E)\right]f_F\,,
\eeq
valid in the sweet region.
The pre-factor of $\frac{1}{4}$ 
in front of the square brackets
is chosen for convenience.
The flavor-indexed quantities $A_\alpha$ and $B_\alpha$, 
$\alpha=e,\mu,\tau$, 
are determined by the neutrino mixing parameters at the Sun's core 
(where $\nu_e\sim \nu_3$) 
and at Earth (vacuum values)~\cite{FN1}.
Since the mixing parameters are energy independent in the sweet region,
so too are $A_\alpha$ and $B_\alpha$.  
The flavor-vectors $B_\alpha$ and $A_\alpha$ 
do depend on the neutrino mass hierarchy 
and on incident neutrino versus antineutrino.
Importantly, 
the entire energy dependence in the evolution equation~\rf{general}
is contained in the level-crossing probability $P_c^h(E)$ 
at the 2nd resonance, given by the expression
\bea{crossprob}
P_c^h(E) & = & \Theta\left(E-E^h_R\right)\;\frac{\exp(-\Gamma\sin^2\theta_{13})-\exp(-\Gamma)}
{1-\exp(-\Gamma)}\nonumber\\
& \simeq & 
\exp(-\Gamma\sin^2\theta_{13})\, .
\eea
In this expression, 
$\Theta$ denotes the unit-step function, 
and ${\cal O}(\delta m^2_{21}/|\delta m^2_{32}|)\sim 0.03$ terms have been dropped. 
As mentioned in the overview section,
the adiabaticity parameter $\Gamma$
may be written as 
\beq{Gamma}
\Gamma = 4\,\pi^2\,\frac{\lambda_\odot}{\lambda_v}\,,
\eeq
where
\beq{lambda_def}
\lambda_\odot\equiv \left|\frac{d}{dr}\ln N_e(r)\right|^{-1}
\eeq
is the scale of density change 
(the distance for an $e$-folding change in the solar density, 
equal to $6.6\times 10^4\,$km)
and $\lambda_v\equiv4\,\pi\,E/\dmsq_{32}$ is the oscillation length in vacuum.
Consequently, 
$\Gamma$ scales as $1/E$, 
and we define a constant $\tildeG$ via 
\beq{tildeG}
\Gamma\equiv\frac{\tildeG}{E}\,,
\eeq
where
\beq{tildeGdef}
\tildeG\simeq 2.5\times
\left( \frac{ |\delta m^2_{32}| }{2.5\times 10^{-3}\,{\rm eV}^2}\right)\,{\rm  TeV}\,,
\eeq
for the higher-energy resonance in solar matter.
With Eqs.~\rf{Gamma}--\rf{tildeGdef} at hand,
it is apparent that 
$\Gamma\gsim 300$. 
This large $\Gamma$ value, 
together with the experimental input $\sin^2\theta_{13}\ll 1$ 
and the fact that $E>E^h_R$ in the sweet region, 
has been used to obtain the second line of Eq.~\rf{crossprob}.

Note that 
the only neutrino mixing parameter
contained in $P_c^h(E)$ is $\theta_{13}$, 
and it enters essentially squared in an exponent.
It follows then, 
that the dominant energy dependence of the terrestrial flavor ratios 
in the sweet region is governed by $\theta_{13}$, 
as advertised in the introduction.
Moreover, 
the sensitivity of flavor evolution through the adiabatic-to-nonadiabatic transition
is exponentially sensitive to $\theta_{13}^2$.

From Eqs.~\rf{crossprob} and \rf{tildeG}, 
we infer the onset energy for $\Ena^h$ 
for the adiabatic-to-nonadiabatic transition to be
\bea{onsetE}
\Ena^h & \equiv & \third\,\tildeG\,\sin^2\theta_{13}\nonumber\\
& = & 800\,\sin^2\theta_{13}\,\left( \frac{ |\delta m^2_{32}| }{2.5\times 10^{-3}\,{\rm eV}^2}\right)\,{\rm GeV}\,.
\eea
Neutrinos with energy $E_R^h < E \alt \Ena^h$ 
will experience adiabatic level-repulsion at the 2nd resonance,
whereas neutrinos with $E \gg \Ena^h$ 
will experience nonadiabatic level-crossing at the 2nd resonance.
We remark that 
although it is true in the Sun that 
the adiabatic-to-nonadiabatic transition energy $\Ena^h$ 
exceeds the resonant energy $E_R^h$, 
this ordering is not in in general guaranteed.
The ratio of the two energies is 
\beq{ENAonER}
\frac{\Ena^h}{E^h_R} = \frac{2\pi}{3}\frac{\sin^2\theta_{13}}{\cos 2\theta_{13}}
\left[\frac{\sqrt{2}\,G_F\,N_e}{|\frac{d}{dr}\ln N_e|} \approx \sqrt{2}\,G_F\,\lambda_\odot\,N_e(0) \right]\,.
\eeq
Numerically, 
the term in brackets is 1800 for the Sun.
It is this fortuitously large value for the Sun that 
allows the adiabatic-to-nonadiabatic transition 
to probe $\theta_{13}$ values 
all the way down to $\theta_{13}\sim (\frac{2\pi}{3} 1800)^{-\half}\sim 1^\circ$.

We proceed by  
defining the energy $\Ehalf$ 
at which the crossing probability is one half.
From Eqs.~\rf{crossprob} and \rf{tildeG} we readily find that 
\beq{Ehalf}
\Ehalf=\frac{\tildeG\sin^2\theta_{13}}{\ln{2}}= 
3.6\sin^2\theta_{13}\,\left(\frac{|\delta m^2_{32}|}{2.5\!\times\! 10^{-3}\,{\rm eV}^2}\right)\,{\rm TeV}\,.
\eeq
Let us also define the width of the change in crossing probability as $\Delta E=E_+ - E_-$,
where $P^h_c(E_-)= 1/e$ and $P^h_c(E_+)= 1-1/e$.
This yields 
\beq{DeltaE}
\Delta E = (e-1)\, (\ln{2}) \, \Ehalf\simeq1.19\,\Ehalf\,.
\eeq
It is seen that 
$\Ehalf$ as well as $\Delta E$ 
are approximately equal to $4\,\sin^2\theta_{13}\approx 4\,\theta^2_{13}$
in units of TeV.
Both $\Ehalf$ and $\Delta E$ are very sensitive probes of the unknown $\theta_{13}$.
The experimental upper bound $\sin\theta_{13}\le 0.22$ at $3\sigma$
establishes that 
$\theta_{13}\simeq\sin\theta_{13}$ to better than 1\% accuracy.
Accordingly, 
we may invert Eq.~\rf{Ehalf} to obtain  
\beq{theta}
\theta_{13}= 1.0^\circ\,\sqrt{\left(\frac{\Ehalf}{{\rm GeV}}\right) 
\left(\frac{2.5\times 10^{-3}\,{\rm eV}^2}{|\delta m^2_{32}|}\right)}\,.
\eeq
For $\theta_{13}$ in the phenomenologically interesting range 
$0.6^\circ \le \theta_{13} \le 3^\circ$,
Eq.~\rf{Ehalf} tells us that 
$\Ehalf$ lies within the sweet region.
Then, 
the inverse Eq.~\rf{theta} establishes that 
a measurement of the adiabatic-to-nonadiabatic transition energy, 
via changing flavor ratios, 
probes the $\theta_{13}$ mixing angle 
in this very interesting range from $0.6^\circ$ to $3^\circ$.
Equation \rf{theta} is the most important result of this paper.  
We show below that 
the flavor signal is also sensitive to the neutrino mass hierarchy.

The ratio $\Delta E/\Ehalf\simeq 1.19$ is independent of $\theta_{13}$.
Consequently, 
the number of events needed to establish $\Ehalf$, 
and thereby $\theta_{13}$, 
is independent of $\theta_{13}$ or $\Ehalf$.
Of course, 
the necessary event number will depend on 
the magnitude of flavor change over the transition region.  
Since $\Delta E\sim\Ehalf$, 
we have that $\Delta E/E=\Delta \ln E\sim 1$.
From this we infer that 
the relevant event sample must span 
about a natural log in energy about $\Ehalf$ to map out the transition region in detail,
again independent of $\theta_{13}$ or $\Ehalf$.
However, 
to measure the gross change in flavor ratios across the transition, 
an event sample may be drawn from an energy region 
as large as the entire sweet region,
0.3 to $10\,$GeV.
The availability of this wide energy range for the data sample is fortunate, 
for at the relatively low energies of the sweet region, 
large uncertainties are expected 
in the reconstruction of the neutrino energy from the measured charged lepton.

The change in the flavor ratios across the \A2N transition region 
of width $\Delta E$ centered on $\Ehalf$
is the signal.  
The magnitude of the flavor change across the transition region 
is determined by Eq.~\rf{general}
and depends on the parameters $A_\alpha$ and $B_\alpha$.
These parameters in turn depend on the neutrino mixing angles,
the mass hierarchy,
and the particle type (neutrinos vs.\ antineutrinos).
The two hierarchies and the neutrino versus antineutrino type lead 
to four possibilities.
To distinguish among these possibilities, 
we employ the notation $A_\alpha^{\rm NH}$, $\overline{A}{}_\alpha^{\rm IH}$, etc.,
where  the bar denotes the antineutrino case and the superscripts NH and IH 
refer to the normal and inverted hierarchy, 
respectively. 
For our present purposes, 
it is sufficient to employ the ``tribimaximal'' values~\cite{TBM} 
for the two large mixing angles, 
i.e., to set 
$\sin^2\theta_{23}^{\rm TBM}=1/2$ and $\sin^2\theta_{12}^{\rm TBM}=1/3$.
The phenomenological accuracy of this approximation is very good,
as is summarized in Ref.~\cite{PRW07}.

Although we employ the tribimaximal values for $\tatmo$ and $\tsun$ in what follows,
we nevertheless exhibit in Table~\ref{table}, 
for possible future use, 
the first-order corrections to the $A_\alpha$ and $B_\alpha$ 
that result if the neutrino mixing angles 
deviate from their tribimaximal values.
To this end, 
we have denoted the deviations from the tribimaximal case by 
$\delta\theta_{jk}\equiv\theta_{jk}-\theta_{jk}^{\rm TBM}$.
The expressions in Table~\ref{table} result from a calculation   
paralleling that in Appendix~D of Ref.~\cite{lw07}.
The $3\sigma$ experimental uncertainties in the angles are~\cite{data}
\bea{angles}
-10.4^\circ  &\le& \delta\theta_{32} \le  8.6^\circ\, , \nonumber \\
-4.8^\circ  &\le& \delta \theta_{21} \le  4.0^\circ \, , \nonumber \\
0^\circ  &\le& \ \theta_{13} \ \le  12.9^\circ  \,.  
\eea
Consequently, 
future data may mandate deviations from 
the tribimaximal-based $A_\alpha$ and $B_\alpha$ used in this work.

%%%%%%%%%%%%%%%%%%%%%%%%%%%%%%%%%%%%%%%%%%
%%%%%%%%%%%%%%%%%%%%%%%%%%%%%%%%%%%%%%%%%%

\section{Results}
\label{sec:results}
In the first subsection, 
we present results valid to all orders in $f_F$.
In particular, 
we provide figures for the $\tau^+\tau^-$ annihilation mode
generated with the complete energy dependence of $f_F(E)$ 
over an energy range that
includes the sweet region.   
In the second subsection, 
we present analytic formulas 
for the sweet region
valid to order $f_F^2$
and compare (favorably) their validity 
with the more exact figures.

\subsection{Exact results}
\label{subsec:exact}
The explicit expressions for the neutrino and antineutrino flavor ratios 
with the normal hierarchy are
\bea{explicitW_NH}
W_{e    }^{\rm NH} &=& \third +\quarter\,(2-2\,P^h_c)\,f_F  \\
W^{\rm NH} _{\mu/\tau} &=& \third+\quarter\,(-1+[1\pm\sqrt{8}\,\theta_{13}\cos\delta]\,P_c^h)\,f_F \nonumber \\
{\overline W}^{\rm NH} _{e    } &=& \third +\quarter\,(-2)\,f_F \nonumber \\
{\overline W}^{\rm NH} _{\mu/\tau} &=& \third +\quarter\,(1\mp\sqrt{8}\,\theta_{13}\cos\delta)\,f_F \,. \nonumber
\eea
The analogous expressions for the flavor ratios with the inverted hierarchy are
\bea{explicitW_IH}
W_{e    }^{\rm IH}   &=& \third  +0\cdot f_F \\
W^{\rm IH} _{\mu/\tau} &=& \third \pm\quarter\,(\sqrt{8}\,\theta_{13}\,\cos\delta)\,f_F \nonumber \\
{\overline W}^{\rm IH} _{e    } &=& \third +\quarter\,(2-4P^h_c)\,f_F \nonumber \\
{\overline W}^{\rm IH} _{\mu/\tau} &=& \third +\quarter\,
 (-1+[2\mp\sqrt{8}\,\theta_{13}\cos\delta]\,P^h_c)\,f_F \,. \nonumber
\eea
In these equations, 
the upper and lower signs 
refer to the $\mu$ and $\tau$ flavors, 
respectively. 
The resonance terms, 
proportional to $P^h_c$, appear where they should:
in the neutrino sector for the NH, 
and in the antineutrino sector for the IH.
Note also that 
in the nonadiabatic limit, 
i.e., 
with $P^h_c$ set equal to one, 
the 2nd resonance becomes irrelevant 
and the NH and IH cases properly reduce to the same result.
Note also that 
the asymmetry between neutrinos and antineutrinos 
is unsurprising:
the solar medium contains no antimatter 
and in this sense provides a CPT-violating background~\cite{CPT}.

In what follows, 
we will omit the small, explicit $\theta_{13}$ dependence 
from Eqs.~\rf{explicitW_NH} and \rf{explicitW_IH} for brevity, 
but we will of course retain the primary $\theta_{13}$ dependence in $P^h_c$ 
as given by Eq.~\rf{crossprob}. 
With $\theta_{13}$ removed, 
the mixing is strictly tribimaximal, 
with its inherent $\nu_\mu\lrarr\nutau$ interchange symmetry: 
the $\pm$~signs in Eqs.~\rf{explicitW_NH} and \rf{explicitW_IH},
which differentiate $\numu$ and $\nutau$, 
disappear.

%%%%%%%%%%%%%%%%%%%%%%%%%%%%%%%%%%%%%%%%%%%%%%%%%%%%%%%%%%%%%%%%
\begin{figure}
\begin{center}
\includegraphics[width=0.95\hsize]{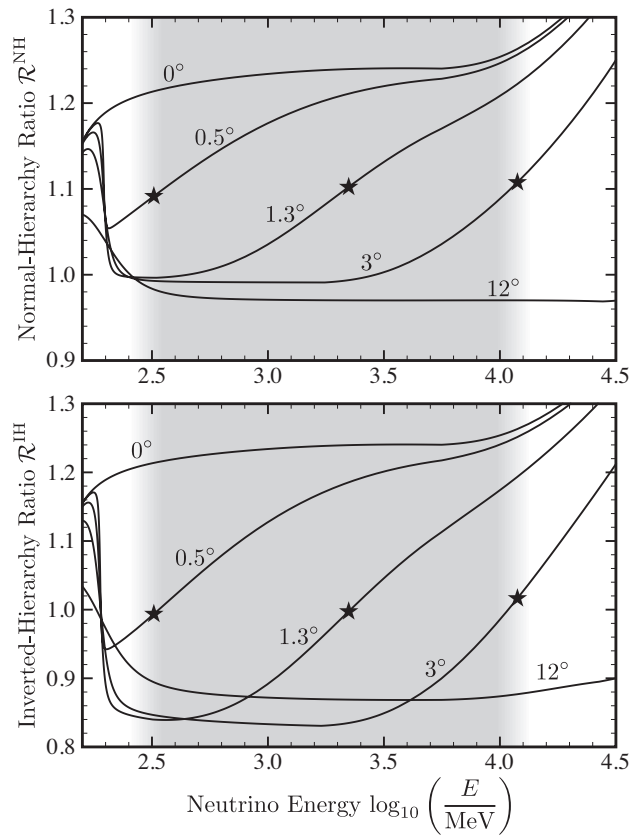}
\end{center}
\vskip-15pt
\caption{The normal-hierarchy ratio ${\cal R}^\textrm{NH}=
(W_\mu^{\rm NH}+{\overline W}_\mu^{\rm NH})/(W_e^{\rm NH}+{\overline W}_e^{\rm NH})$ (top) 
and inverted-hierarchy ratio ${\cal R}^\textrm{IH}=  
(W_\mu^{\rm IH}+{\overline W}_\mu^{\rm IH})/(W_e^{\rm IH}+{\overline W}_e^{\rm IH})$ (bottom) 
versus the neutrino energy $E$ 
for the indicated $\theta_{13}$ mixing angles. 
Here, 
$M_{\rm  WIMP} = 100\,$GeV, 
and the annihilation mode is assumed to be predominantly $\tau^+\tau^-$. 
The approximate location of the ``sweet region,''
where our analytical results apply, 
is shaded. 
The star ($\star$) indicates the location of the energy $E_{1/2}$ 
at which the crossing probability is one half.
The neutrino mass-squared differences are 
$\delta m^2_{21}=8.0\times10^{-5}\,$eV${}^2$ 
and $\delta m^2_{32}=\pm3.0\times10^{-3}\,$eV${}^2$,
$\theta_{21}$ and $\theta_{32}$ are given their tribimaximal values, 
and the CP-violating phase $\delta$ is set to zero.
The family of curves are quite sensitive to the value of $\theta_{13}$, 
even below a degree.
Relative changes in ${\cal R}$ are 20\% for the normal hierarchy (top) 
and 30\% for the inverted hierarchy (bottom), 
with the centered energy $\Ehalf$ 
showing quadratic sensitivity to $\theta_{13}$ 
in agreement with Eq.~\rf{Ehalf}.
We remark that 
the nontrivial behavior 
just below the sweet region
results primarily from the lower-energy resonance
as well as from the energy dependence of $U_m(E)$.
}
\label{fig1} 
\end{figure} 
%%%%%%%%%%%%%%%%%%%%%%%%%%%%%%%%%%%%%%%%%%%%%%%%%%%%%%%%%%%%%%%%%%%%%%

We now turn to the event types in the detector.
A minority of the events will be neutral current (NC) events.
NC events yield no flavor information, 
and we ignore them.
The majority of events will be charged-current (CC) events 
producing a charged lepton.
Among these CC events, 
we note that 
the threshold in energy for production of a charged $\tau$ is 
$m_\tau\,(1+m_\tau/2\,m_N)=3.47\,$GeV.  
Since this value falls in the middle of the sweet region,
it presents a kinematic complication.
It is therefore advantageous to consider observables that 
do not depend on CC $\tau$ production.
For his reason, 
we examine the ratios ($W_\mu/W_e$), 
(${\overline W}_\mu/{\overline W}_e$), 
(${\overline W}_e/W_e$), 
and  (${\overline W}_\mu/W_\mu$).  
The threshold for the muon CC is $E=m_\mu\,(1+m_\mu/2\,m_N)=110\,$MeV,
well below the lower end ($300\,$MeV) of our sweet region.
For experiments with charge identification, 
e.g., 
the proposed large magnetized iron INO experiment,
these ratios may be inferred from measurements of $\numu$ CC events, 
which necessarily contain a muon-track signature, 
as well as from $\nue$ CC events, 
which necessarily contain an electromagnetic plus hadronic shower 
and no muon track.
For experiments without charge identification, 
which can contain a much larger target mass,
we must sum $\nu$ and $\bar\nu$ and work with the single ratio 
${\cal R}\equiv (W_\mu+{\overline W}_\mu)/(W_e+{\overline W}_e)$.

In practice, 
there will be experimental efficiencies 
that differ significantly for the detection of neutrinos versus antineutrinos, 
and muon versus electron CC reactions.
Among other sources of these efficiency differences 
are the unequal CC cross sections 
for scattering of $\numu$ versus $\nu_e$ versus $\numubar$ versus $\nuebar$.
Over the energy range of the sweet region, 
the nature of neutrino scattering is quite energy dependent.
Relevant cross sections include inverse $\beta$~decay,
neutrino--electron scattering, 
quasi-elastic scattering, 
pion production, 
and deep inelastic scattering. 
We note that 
efficiency differences may be advantageous 
if they can mitigate the mixing of 
$\numu$, $\nu_e$, $\numubar$, and $\nuebar$ signals.
However, 
in this work we will simply assume uniform efficiencies for detection of 
$\numu$, $\nu_e$, $\numubar$, and $\nuebar$,
rather than introduce another layer of complexity.

%%%%%%%%%%%%%%%%%%%%%%%%%%%%%%%%%%%%%%%%%%%%%%%%%%%%%%%%%%%%%%%%%%%%%%%%%%%%
\begin{figure}
\begin{center}
\includegraphics[width=0.95\hsize]{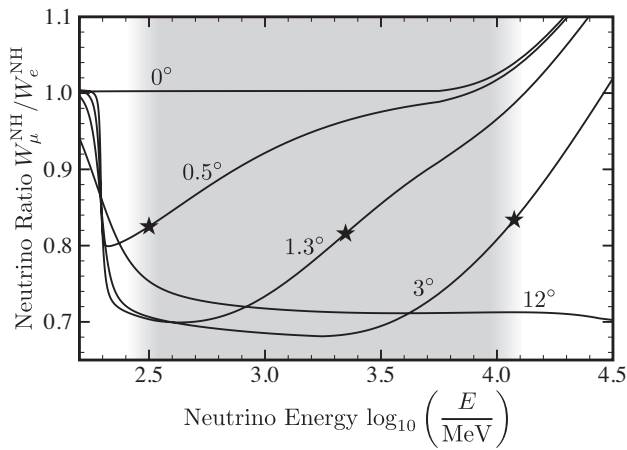}
\end{center}
\vskip-15pt
\caption{The normal-hierarchy ratio $W^\textrm{NH}_\mu/W^\textrm{NH}_e$ 
versus the neutrino energy $E$ 
for various $\theta_{13}$ mixing angles. 
Input parameters are identical to those in Fig.~\ref{fig1}.
The relative change in flavor ratio is about 40\% across the transition region.
Again, 
$\Ehalf$ (denoted by $\star$) 
depends quadratically on $\theta_{13}$,
allowing a potential inference of this mixing angle.
In the normal hierarchy,
the 2nd resonance occurs in the neutrino sector 
and not in the antineutrino sector.
}
\label{fig2} 
\end{figure} 
%%%%%%%%%%%%%%%%%%%%%%%%%%%%%%%%%%%%%%%%%%%%%%%%

%%%%%%%%%%%%%%%%%%%%%%%%%%%
\begin{figure}
\begin{center}
\includegraphics[width=0.95\hsize]{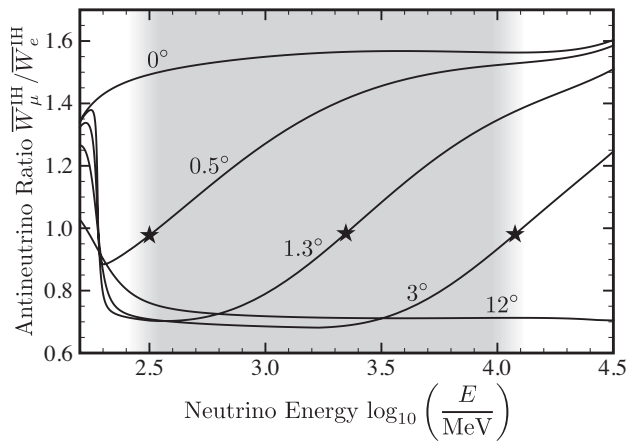}
\end{center}
\vskip-15pt
\caption{The inverted-hierarchy ratio 
$\overline{W}^\textrm{IH}_\mu/\overline{W}^\textrm{IH}_e$
versus the neutrino energy $E$ 
for various $\theta_{13}$ mixing angles. 
The input parameters are identical to those in Fig.~\ref{fig1}. 
The relative change in flavor ratio is almost a factor of two across the transition region.
Again, 
$\Ehalf$ (denoted by $\star$) depends quadratically on $\theta_{13}$
allowing a potential deduction of this mixing angle.
In the inverted hierarchy,
the 2nd resonance occurs in the antineutrino sector 
and not in the neutrino sector.
}
\label{fig3} 
\end{figure} 
%%%%%%%%%%%%%%%%%%%%%%%%%%%%%%%%%%%%%%%%%%%%%%%%%%%%%%%%%%%%%%%%%%%%%%%%%

The ratio ${\cal R}=(W_\mu+{\overline W}_\mu)/(W_e+{\overline W}_e)$ is defined 
to weight equally the event numbers of neutrino and antineutrinos.
Since the 2nd resonance occurs only in the neutrino sector (NH) 
or only in the antineutrino sector (IH), 
but not in both,
the contribution of the non-resonant sector (neutrino or antineutrino) 
will mitigate the contribution from the resonant sector.
In this sense, 
${\cal R}$ as defined is a conservative variable.

The energy dependence of the ratio ${\cal R}$ 
is presented without approximation in Fig.~\ref{fig1}. 
One sees that even for this conservative variable, 
20--30\% changes in the relative flavor ratio occur 
across the \A2N transition in the $0.3$--$10\,$GeV sweet region.
Curves are parameterized by values of $\theta_{13}$,
with significant effects apparent all the way down to a fraction of a degree. 
The quadratic dependence of $\Ehalf$
(defined as the energy where the nonadiabatic crossing probability is 50\%
and denoted in the figures by a $\star$) 
on $\theta_{13}$
is clearly seen.
In favorable circumstances, 
this ultra-sensitivity of the spectrum to $\theta_{13}$ 
could potentially be used to measure~$\theta_{13}$.

Notable is that 
the change in the flavor ratio 
is about twice as large for the inverted neutrino mass hierarchy 
as it is for the normal hierarchy.  
We can dissect some of this behavior by 
turning to ratios defined only in the neutrino sector 
or only in the antineutrino sector.
By excluding the non-resonant sector from the flavor ratio,
we expect the change in the ratio across the \A2N transition 
to be about twice that of the sector-summed result.
That is, 
we expect $\sim 40\%$ effects in the neutrino sector 
if the hierarchy is normal, 
and $\sim 80\%$ effects in the antineutrino sector 
if the hierarchy is inverted.
In Figs.~\ref{fig2} and \ref{fig3}, 
we see that 
this expectation is met.
Consequently, 
if some experimental handle is available 
to even partially separate the neutrino and antineutrino data samples, 
then considerably greater discriminatory power becomes available.
Put another way, 
across the sweet region, 
with the normal hierarchy, 
the neutrino flavor ratio should exhibit 
considerable energy dependence 
and the antineutrino flavor ratio should not. 
Conversely, 
with the inverted hierarchy, 
the antineutrino flavor ratio should show 
significant energy dependence 
and the neutrino flavor ratio should not. 
Thus, 
if neutrino--antineutrino discrimination becomes possible, 
then the observed energy dependence of neutrino versus antineutrino flavor ratios differentiates between the two possible mass hierarchies.

\subsection{Approximate results}
\label{subsec:approx}
With the exact evolved flavor relations of Eqs.~\rf{explicitW_NH} and \rf{explicitW_IH} 
in hand, 
we may take ratios and expand in powers of the small $f_F$'s 
given in Table~\ref{fF_table}.
To order $f_F^2$, 
one finds for the normal hierarchy:
\bea{mu2eNH}
\label{ANH}
{\cal R}^{\rm NH} & = & 
1+\frac{9}{8}\,P^h_c\,f_F + \frac{27}{32}\,(P^h_c)^2\,f_F^2\,,\\
\label{CNH}
\left( \frac{W^{\rm NH}_\mu}{W^{\rm NH}_e}\right) &=&
  1-\frac{9}{4}\,(1-P^h_c)\,f_F + \frac{27}{8}(1-P^h_c)^2\,f_F^2\,, \\
\label{DNH}
\left( \frac{{\overline W}^{\rm NH}_\mu}{{\overline W}^{\rm NH}_e}\right) & =& 
  1+\frac{9}{4}\,f_F + \frac{27}{8}\,f_F^2\,, \\
\label{ENH}
\left( \frac{{\overline W}^{\rm NH}_e}{W^{\rm NH}_e}\right) & =& 
 1-\frac{3}{2}\,(2-P^h_c)\,f_F + \frac{9}{4}P^h_c\,(1-P^h_c)\,f_F^2\,, \\
\label{FNH}
\left( \frac{{\overline W}^{\rm NH}_\mu}{W^{\rm NH}_\mu}\right) & =&
 1-\frac{3}{4}\,(2-P^h_c)\,f_F + \frac{9}{16}P^h_c\,(1-P^h_c)\,f_F^2\,. \hspace{8mm}
\eea
These equations reflect the fact that 
with the normal hierarchy, 
the 2nd resonance and the \A2N transition lie in the neutrino sector
and not in the antineutrino sector.

The analogous calculation for the inverted hierarchy yields
\bea{mu2eIH}
\label{AIH}
{\cal R}^{\rm IH} & = & 
1-\frac{9}{8}\,(1-2P^h_c)\,f_F -\frac{9}{32}\,(1-2P^h_c)^2\,f_F^2\,,\hspace{6mm}\\
\label{CIH}
\left( \frac{W^{\rm IH}_\mu}{W^{\rm IH}_e}\right) & = &  
  1+0\cdot f_F + 0\cdot f_F^2\,, \\
\label{DIH}
\left( \frac{{\overline W}^{\rm IH}_\mu}{{\overline W}^{\rm IH}_e}\right) & = & 
  1-\frac{9}{4}\,(1-2P^h_c)\,f_F + \frac{27}{8}(1-2P^h_c)^2\,f_F^2\,, \\
\label{EIH}
\left( \frac{{\overline W}^{\rm IH}_e}{W^{\rm IH}_e}\right) & = &  
  1+\frac{3}{2}\,(1-2P^h_c)\,f_F + 0\cdot f_F^2\,, \\
\label{FIH}
\left( \frac{{\overline W}^{\rm IH}_\mu}{W^{\rm IH}_\mu}\right) & = &  
  1-\frac{3}{4}\,(1-2P^h_c)\,f_F +  0\cdot f_F^2 \,.
\eea
The equations here reflect the fact that 
with the inverted hierarchy, 
the 2nd resonance and the \A2N transition lie in the antineutrino sector
and not in the neutrino sector.

For comparison, 
we remind the reader that 
in the absence of matter, 
the analogous vacuum-evolved ratios are 
$(W_\mu/W_e)_{\rm vacuum}=
{(\overline{W}}_\mu/{\overline{W}}_e)_{\rm vacuum} = 1+\frac{3}{2}\,f_F+f_F^2$, 
as has been established in Sec.~\ref{sec:theory}.
For the value $f_F=0.18$ 
appropriate to the $\tau^+\tau^-$ annihilation mode in the sweet region,
this ratio is equal to 1.30.
For the value $f_F=-0.09$ 
appropriate to the $b\,\bbar$ annihilation mode in the sweet region,
this ratio is equal to 0.9.
Inserting either of these two $f_F$ values 
into the ratios calculated with solar-matter effects,
Eqs.~\rf{ANH}--\rf{FIH}, 
gives very different results.

We note that 
the change in flavor ratios over the transition from adiabatic ($P^h_c=0$)
to nonadiabatic ($P^h_c\to\cos^2\theta_{13}\simeq1$)
is twice as large for the antineutrino sector in the inverted hierarchy,
as compared to the neutrino sector in the normal hierarchy.
This factor of two is subtle.
We reveal its origin in the Appendix.

We may now use Eqs.~\rf{ANH}--\rf{FIH} 
to obtain the magnitude of changes in the flavor ratios 
as the neutrino energy $E$ varies across the sweet region. 
The energy-dependent $P_c^h(E)$ is a monotonic function  
and obeys $P_c^h(E\to 0)=0$ and $P_c^h(E\to \infty)=\cos^2\theta_{13}$. 
Thus, 
at leading order in $\theta_{13}$ 
the range of the crossing probability 
is $0\lsim P_c^h(E)\lsim 1$. 
We here list a sample of ratios of the quantities in Eqs.~\rf{ANH}--\rf{FIH} 
above and below the \A2N transition,
calculated to order $f_F^2$, 
and then evaluated with the choices $f_F = +0.18$ 
for the $\tau^+\tau^-$ mode
and $f_F = -0.09$ 
for the $b\,\bbar$ mode.
In the normal hierarchy we have
\beq{mu2echange1}
\frac{{\cal R}^{\rm NH}_{P=1}}{R^{\rm NH}_{P=0}} = 1+\frac{9}{8}\,f_F +\frac{27}{32}\,f_F^2 
= \left\{ 
\begin{array}{lc}
  1+0.23 & (\tau^+\tau^-) \\ 
  1-0.09 & (b\,\bbar) \\ 
\end{array} \right. ,
\eeq
\beq{mu2echange2}
\frac{(W^{\rm NH}_\mu/W^{\rm NH}_e)_{P=1}}{(W^{\rm NH}_\mu/W^{\rm NH}_e)_{P=0}} = 
1+\frac{9}{4}\,f_F + \frac{27}{16}\,f_F^2 
= \left\{ 
\begin{array}{lc}
  1+0.46 & \\ 
  1-0.19 & \\ 
\end{array} \right. ,
\eeq
while in the inverted hierarchy we have
\beq{mu2echange3}
\frac{{\cal R}^{\rm IH}_{P=1}}{R^{\rm IH}_{P=0}} = 1+\frac{9}{4}\,f_F +\frac{189}{64}\,f_F^2
= \left\{ 
\begin{array}{lc}
  1+0.50 & (\tau^+\tau^-) \\
  1-0.18 & (b\,\bbar) \\
\end{array} \right. ,
\eeq
\beq{mu2echange4}
\frac{ ({\overline W}^{\rm IH}_\mu / {\overline W}^{\rm IH}_e)_{P=1} }
{ ({\overline W}^{\rm IH}_\mu / {\overline W}^{\rm IH}_e)_{P=0} } =
1+\frac{9}{2}\,f_F +\frac{81}{8}\,f_F^2 
= \left\{ 
\begin{array}{lc}
  1+1.14 & \\ 
  1-0.32 & \\
\end{array} \right. .
\eeq
The value of these ratios of ratios in vacuum is unity, 
of course.
The large flavor-ratio differences here 
are due to the \A2N transition at the matter-induced 2nd resonance.
For the normal hierarchy, 
the ratio of ratios for the neutrino changes by 23\% and 46\% 
in the $\tau^+\tau^-$ annihilation mode, 
and by 10\% and 20\% 
in the $b\,\bbar$ mode.
For the inverted hierarchy, 
the ratio of ratios for the antineutrino changes by slightly more than twice that,
by 50\% and 114\% in the $\tau^+\tau^-$ annihilation mode, 
and by 20\% and 32\% in the $b\,\bbar$ mode.
The changes for the two hierarchies are of opposite sign.
Together with the very little change expected in the $W^+W^-$ annihilation mode,
one sees that just an inference of the existence and sign of the change 
can discriminate among 
the three most popular annihilation modes of solar WIMPs.

The qualitative factors for the change in the ratio of ratios, 
given in Eqs.~\rf{mu2echange1}--\rf{mu2echange4},
are borne out in the more exact results 
shown in Figs.~\ref{fig1}--\ref{fig3}.

\section{Discussion and conclusion}
\label{sec:discon}
We have shown that \\
$\bullet$
For neutrinos from the $\tau^+\tau^-$ annihilation mode of solar WIMPs,
10--100\% changes in neutrino flavor ratios at Earth are expected 
in the $0.3$--$10\,$GeV energy interval.
This energy interval is a mapping 
across the \A2N transition of the higher-energy 2nd resonance in the Sun.
Whether the change is 10\% or 100\%, 
or in between, 
depends on which flavor-ratio observable is available to the experimenter.\\
$\bullet$ 
For the $b\,\bbar$ annihilation mode of the WIMPs, 
the flavor change is about half as large,
and of opposite sign. \\ 
$\bullet$
For the $W^+W^-$ annihilation mode, 
the flavor change is small.\\
$\bullet$
Arising from the higher-energy resonance in the Sun, the effect 
occurs in the neutrino sector but not antineutrino sector if 
the neutrino mass hierarchy is normal,
and in the antineutrino but not neutrino sector if the mass hierarchy is inverted.\\
$\bullet$
Thus, 
a measurement of the sign and magnitude of this flavor change 
could indicate the WIMP annihilation mode, 
and a determination of the resonant sector, 
neutrino or antineutrino, 
would indicate the mass hierarchy.\\

A plot of flavor versus energy determines the energy $\Ehalf$ 
at which the \A2N transition probability is 50\%
and determines the width (in energy) $\Delta E$ of the transition.
It turns out that\\
$\bullet$
$\Ehalf\approx\Delta E$, 
and each is proportional to $\theta_{13}^2$.
Thus, 
the system is over-constrained, 
which provides a clean signal for the underlying physics.\\
$\bullet$
Moreover, 
the quadratic dependence on $\theta_{13}$ 
provides a very sensitive probe for measuring the value of $\theta_{13}$. \\
$\bullet$
The \A2N transition profile is measurable for $\theta_{13}$ as small as $0.5^\circ$.\\

To establish experimentally 
a 30\% change in the flavor ratios over the transition region 
with an $n$-sigma significance,
the number of required CC $\nue$ and $\numu$ events is $\sim 2\,(n/30\%)^2$.
(The pre-factor 2 arises from the need for statistical accuracy 
on each side of the transition.)
Thus, 
for $3\sigma$ significance 
one needs roughly 200 solar WIMP events 
separated from the atmospheric background
and spanning the sweet region between $0.3$ and $10\,$GeV.
According to Eq.~\rf{fidrate}, 
a 200-event sample 
can be accumulated in the sweet region 
with a megaton detector in 1.5~years,
if the WIMP annihilation cross section saturates its present experimental upper limit 
of $3\times 10^{-38}{\rm cm}^2$.
The annihilation cross section could have this value, 
or it could be orders of magnitude less.

We end with mention of the many mega-detectors 
for low-energy neutrinos that have been proposed.
They represent a presence of hope and vision that will shape the future.
Proposed detector materials include water for Cerenkov signals 
(MEMPHYS, UNO, and HyperK),
scintillators such as liquid argon 
(LENA, GLACIER, LArTPC, and TASD),
and magnetic iron calorimeters 
(MIND, MONOLITH, and INO).
The magnetized detectors can distinguish neutrino and antineutrino events,
which enhances the signal explored herein by a factor of two,
and thereby reduces the data required by a factor of four.
But magnetized detectors cannot be built as large as unmagnetized detectors.

\acknowledgments

T.J.W.~thanks Nicole Bell and the University of Melbourne for hospitality and support
as well as Fido's Coffeehouse in Nashville, TN 
for a charming caffeine-rich writing environment.
This work is supported in part by the DOE 
under Grant No.\ DE-FG05-85ER40226, 
by the European Commission
under Grant No.\ MOIF-CT-2005-008687,
and by CONACyT under Grant No.\ 55310.

\appendix*

\section{The factor of two between antineutrino and neutrino adiabatic-to-nonadiabatic evolution}
\label{app:one}
To explain the relative factor of two 
at order $f_F$ 
between the neutrino's and the antineutrino's \A2N evolution, 
we use the density-matrix formalism.
In the flavor basis, 
the density matrix at Sun's core is 
$\rho_F=\third\,\openone+f_F\,\left[\,2\,|\nutau\ketbra\nutau|-|\numu\ketbra\numu|-|\nue\ketbra\nue|\,\right]$.
Invoking 
the completeness relation $\sum_\alpha |\nualpha\ketbra\nualpha|=\openone$ 
as well as maximal neutrino mixing to symmetrize
$\left(\,2\,|\nutau\ketbra\nutau|-|\numu\ketbra\numu|\,\right)
\rarr \half\left(\,|\nutau\ketbra\nutau|+|\numu\ketbra\numu|\,\right)$
shows that this matrix may be rewritten in terms of just the $\nue$ projector as 
$\rho_F\rarr \third\left[\,(1+\frac{3}{2}\,f_F)\,\openone -\frac{9}{2}\,f_F\,|\nue\ketbra\nue|\,\right]$.

Now, 
with the normal hierarchy, 
and in the lower-energy adiabatic region, 
the $\nue$ emerges from the Sun as $\nu_3$, 
and the $\numu$ and $\nutau$ are left to emerge as $\nu_1$ and $\nu_2$.
In the higher-energy nonadiabatic region, 
the $\nue$ emerges as $\nu_2$, 
while the $\numu$ and $\nutau$ emerge as $\nu_1$ and $\nu_3$.
(Recall that we have chosen the sweet region 
to be below the nonadiabatic onset $E^l_{NA}$ of the lower-energy 1st resonance, 
and so the conversion across the 1st resonance is $\nue \leftrightarrow \nu_2$.)

On the other hand, 
with the inverted hierarchy, 
and in the lower-energy adiabatic region, 
the $\nuebar$ again emerges as ${\bar \nu}_3$ 
and the $\numubar$ and $\nutaubar$ emerge as ${\bar \nu}_1$ and ${\bar \nu}_2$. 
However, 
in the higher-energy nonadiabatic region 
the $\nuebar$ emerges as ${\bar \nu}_1$,
while the $\numubar$ and $\nutaubar$ 
are left to emerge as ${\bar \nu}_1$ and~${\bar \nu}_2$.

Thus, 
at lower energies we have $\nue\rarr\nu_3$ and $\nuebar\rarr {\bar \nu}_3$ 
in the normal and inverted hierarchies, 
respectively, 
while at higher energies we have 
$\nue\rarr\nu_2$ and $\nuebar\rarr {\bar \nu}_1$ 
in the normal and inverted hierarchies, 
respectively.
Replacing the $\nue$ projector $|\nue\ketbra\nue|$ in $\rho_F$
with the appropriately evolved $\nu$-mass projector 
$|\nu_1\ketbra\nu_1|$ or $|\nu_2\ketbra\nu_2|$ or $|\nu_3\ketbra\nu_3|$, 
and then calculating the final flavor-matrix elements 
$W_\mu=\langle\numu| \rho_F |\numu\rangle$ 
and $W_e=\langle\nue| \rho_F |\nue\rangle$,
one readily obtains the change in flavor ratios $(W_\mu/W_e)$ 
across the adiabatic-to-nonadiabatic transition.
To order $f_F$, 
the result is $\frac{9}{2}\,f_F\,(\U2_{e2}-\U2_{\mu 2}+\U2_{\mu 3}-\U2_{e 3})$ 
for the normal hierarchy,
and $\frac{9}{2}\,f_F\,(\U2_{e1}-\U2_{\mu 1}+\U2_{\mu 3}-\U2_{e 3})$ 
for the inverted hierarchy.
Putting into these expressions the values of tribimaximal mixing, 
one finds that the change is $\frac{9}{4}\,f_F$ 
for the neutrino sector in the normal hierarchy, 
and twice that for the antineutrino sector in the inverted hierarchy.
Notice that this result requires the mixing values of $\nue$ and $\numu$ 
with all three neutrino mass states.
In the end, 
it is a ``conspiracy'' of the tribimaximal values that 
gives the seemingly simple factor of 2.

%%%%%%%%%%%%%%%%%%%%%%%%%%%%%%%%%%%%%%%%%%%%%%%%%%%%%%%%%%%%%%%%%%%%%%%%%%%%%%%%%%%%%%%%%%%%%%%%%%%%%%%%%%%%%%%%%%%%%%%


\begin{thebibliography}{99}
\bibitem{comment}
We disregard the possibility of tau-neutrino regeneration, 
which would absorb high-energy neutrinos 
and re-inject them into the flux at lower energies. 
See also Footnote~\cite{fn}.
\bibitem{fn}
For very heavy WIMP masses, 
the energy independence of flavor ratios 
will be affected by $\nutau$ ``regeneration.''
Neutrinos with energy above $\sim 100\,$GeV
interact with the solar matter
(the absorption probability is 60\% at $M_\textrm{WIMP}\simeq200\,$GeV).
Via the CC interaction, 
the $\nutau$'s produce $\tau$'s. 
Each $\tau$ promptly decays to produce a lower-energy $\nutau$.
The cycle repeats, 
leading to a lower-energy flux of neutrinos, 
mostly $\nutau$'s.
In general, 
this will contribute  
a calculable ``secondary'' $E$-dependence 
to the neutrino flavors in the sweet region.
\bibitem{WWnot}
The erratum in Ref.~\cite{Cirelli05} 
corrects a programming error in their earlier paper that
had suggested that 
the neutrino flavor ratio coming from the $W^+W^-$ mode 
was energy independent, too.
The earlier, 
erroneous energy-independent flavor ratio for $W^+W^-$ 
was transcribed into Fig.~6 of Ref.~\cite{lw07}.
\bibitem{absorpnote} 
We note that neutrinos are not appreciably absorbed in the Sun 
until their energies reach $50\,$GeV and above~\cite{lw07}.
\bibitem{spin-indep}
We note that 
present limits on the spin-independent cross section of the WIMP 
are five orders of magnitude below that 
of the spin-dependent cross section.
Such a huge ratio of the two cross-section types is a theoretical possibility,
as this ratio depends sensitively on the nature of the WIMP.
\bibitem{pi_note}
One might hope that 
the decay of the many pions produced in WIMP annihilation 
would enhance the flux of neutrinos at lower energies.  
However, 
the density of nucleons in the Sun's core is $n_N\sim 70\,N_A/{\rm cm^3}$,
and the pion interaction cross section is strong, $\sigma_{\pi N}\sim 40\,$mb. 
It follows that 
the mean interaction time is 
$t_{\rm int}=(c\,n_N\,\sigma_{\pi N})^{-1}\sim 2\times 10^{-11}\,$s.
This is to be compared to the charged-pion lifetime 
of $\tau_{\pi^\pm}=(E_\pi /m_\pi)\times 2\times 10^{-7}\,$s.
Clearly, 
the pion loses energy and becomes innocuous 
through many subsequent absorptions and re-emissions,
long before it can decay.
\bibitem{FN1}
In the language of the matrices  $\U2$, $\P$, and $\U2_{\hspace{-.3mm}m}^T(0)$,
where $\P$ is spelled out in Ref.~\cite{lw07},
the flavor vectors can be expressed as follows: 
$A_\alpha= 4\,\U2\,\U2_{\hspace{-.3mm}m}^T (0)\,(-1,-1,+2)^T$
and 
$B_\alpha=4\,\U2\,(\P-\openone)\,\U2_{\hspace{-.3mm}m}^T (0)\,(-1,-1,+2)^T$.
\end{thebibliography}
\end{document}